\documentclass[referee,pdflatex,sn-mathphys-ay]{sn-jnl}

\usepackage{graphicx}
\usepackage{multirow}
\usepackage{amsmath,amssymb,amsfonts}
\usepackage{amsthm}
\usepackage{mathrsfs}
\usepackage[title]{appendix}
\usepackage{xcolor}
\usepackage{textcomp}
\usepackage{manyfoot}
\usepackage{booktabs}
\usepackage{algorithm}
\usepackage{algorithmicx}
\usepackage{algpseudocode}
\usepackage{listings}

\theoremstyle{thmstyleone}

\theoremstyle{thmstyletwo}

\theoremstyle{thmstylethree}

\makeatletter
\@ifundefined{th@thmstylethree}{\theoremstyle{definition}}{\theoremstyle{thmstylethree}}
\makeatother
\newtheorem{criterion}{Falsification criterion}

\begin{document}

\title[A Distinct Communication Strategies Model of the Double Empathy Problem]{A Distinct Communication Strategies Model of the Double Empathy Problem}

\author*[1]{\fnm{Enrique} \sur{Calderoli}}\email{enrique\_calderoli@hotmail.com}

\author[2]{\fnm{Tiago} \sur{Figueiredo}}\email{tiagofigueiredosf@gmail.com}

\author[3]{\fnm{Maria Cristina} \sur{Varriale}}\email{cris@mat.ufrgs.br}

\author[1,4,5,6]{\fnm{Flávio} \sur{Kapczinski}}\email{fkapczinski@hcpa.edu.br}

\affil[1]{\orgdiv{Department of Psychiatry}, \orgname{Universidade Federal do Rio Grande do Sul (UFRGS)}, \orgaddress{\street{Ramiro Barcelos, 2400}, \city{Porto Alegre}, \postcode{90035-903}, \state{Rio Grande do Sul}, \country{Brazil}}}

\affil[2]{\orgname{Instituto COGNUS de Ensino e Pesquisa – ICEP - Centro Clínico Advance I}, \orgaddress{\street{Q SGAS, 915}, \city{Brasília}, \postcode{70390-150}, \state{Distrito Federal}, \country{Brazil}}}

\affil[3]{\orgdiv{Institute of Mathematics and Statistics}, \orgname{Universidade Federal do Rio Grande do Sul (UFRGS)}, \orgaddress{\street{Av. Bento Gonçalves, 9500}, \city{Porto Alegre}, \postcode{91509-900}, \state{Rio Grande do Sul}, \country{Brazil}}}

\affil[4]{\orgdiv{Department of Psychiatry and Behavioural Neurosciences}, \orgname{McMaster University}, \orgaddress{\street{100 West 5th Street}, \city{Hamilton}, \postcode{L8N 3K7}, \state{Ontario}, \country{Canada}}}

\affil[5]{\orgdiv{Mood Disorder Program}, \orgname{St. Joseph’s Healthcare Hamilton}, \orgaddress{\street{100 West 5th Street}, \city{Hamilton}, \postcode{L8N 3K7}, \state{Ontario}, \country{Canada}}}

\affil[6]{\orgdiv{\orgname{Instituto Nacional de Ciência e Tecnologia Translacional em Medicina (INCT-TM)}, \orgaddress{\street{Ramiro Barcelos, 2350}, \city{Porto Alegre}, \postcode{90035-903}, \state{Rio Grande do Sul}, \country{Brazil}}}}

\abstract{The double empathy problem reframes social difficulties between autistic and non-autistic people as reciprocal rather than as deficits located exclusively within the autistic individual. While there has been ample discussion about communicative compatibility, interaction cost, and affective resonance in cross-neurotype interactions, no mechanistic model of how these processes interact has been proposed. Here we build a feedback-loop mathematical model that would theoretically reproduce the empathy degradation observed during communication in neurotypical-autistic pairs solely due to differences in communication preferences of each neurotype. Numerical simulations of dyadic interactions show the model, whose mechanism is based solely on channel preferences, can illustrate the breakdown of empathic bonding observed clinically. This mathematical model presents testable predictions regarding the verbal dimension of dyadic empathic bonding and represents a promising step towards a formal account of a broader framework of social interaction dynamics of neurodivergent individuals and possible mitigation strategies for the double empathy problem.}


\maketitle

\section{Introduction}\label{sec:introduction}

Autistic social communication was historically interpreted through deficit-centered models that located difficulty primarily within the autistic person. Such accounts often treated non-autistic communicative conventions as a neutral benchmark and explained misunderstanding through impaired social cognition, reduced empathy, or atypical use of verbal and nonverbal cues \citep{baroncohen1985theory,baroncohen1995mindblindness,baroncohen2002extreme}. Although this literature identified important individual differences, it frequently underrepresented the relational character of social understanding.  

The double empathy problem shifted this perspective by proposing that misunderstanding between autistic and non-autistic people is bidirectional and emerges from differences in lived experience, communication, and expectations \citep{milton2012ontological,milton2022ten}. Findings of effective autistic-to-autistic information transfer, greater rapport in neurotype-matched dyads, and difficulty among non-autistic observers when interpreting autistic people are consistent with the need to study the fit between partners rather than locating failure in either individual alone \citep{crompton2020information,crompton2020neurotype,edey2016interaction,sheppard2016read,sasson2017thin}. 

Autistic communication is heterogeneous, and neither autistic nor non-autistic people can be perfectly represented by a single fixed profile. Group-level differences have nevertheless been reported in literal language processing \citep{vulchanova2015figurative}, prosody \citep{shriberg2001speech,mccann2003prosody}, nonverbal signaling, and the interpretation of ambiguous or implicit cues. These differences may alter how much effort each partner must invest to infer intention, predict responses, and coordinate behavior, as well as their ability to properly capture and weight all signals emitted by the other partner. The relevant variable is therefore not a communication feature considered in isolation, but the compatibility between the expressive and interpretive repertoires of both partners. 

We propose that the double empathy problem can be more precisely understood as a dynamic problem of dyadic social coupling, with communicative mismatch as one of its main dimensions. The canonical (though not universal) communicative profiles of autistic and neurotypical individuals – preference for verbal and literal communication and lack of preference between verbal and nonverbal channels, respectively –, can significantly increase interaction cost in cross-neurotype dyads. The model developed here provides a formal account of the channel preference mismatch contributing to the double empathy problem. Combined with other relevant dimensions of empathic bonding, such as affective resonance, accommodation, persistence, cognitive and emotional fatigue, it represents the beginning of a possible unified model of emotional communication and bonding in dyads. 

\section{Methods}\label{sec:methods}

In mathematics, functions describe mappings $f: X\to Y$ between elements of sets $X$ and $Y$. Neuropsychological functions are the analogous objects used to describe how different neurological or mental quantities are connected in explicit or implicit human behavior and cognition. In precise terms, they are defined by measurable mappings between states of internal or external quantities and latent states or overt behaviors. The parameters these functions take represent cognitive characteristics of the individual whose cognition or behavior is being modeled. These maps prove that emotional states are not undefinable concepts beyond the reaches of quantitative science, but clear manifestations of cognitive processes that can be modeled biologically and mathematically and generate testable predictions \citep{ferrer2013dynamical,gottman2002mathematics,treur2013integrative}. The combination of several neuropsychological functions defines a dynamical system, either in discrete time or continuous time \citep{hoeksma2004emotion,lewis2005bridging}. 

The canonical framework for employing these functions to model human emotions employs a dynamical systems approach \citep{lewis2005bridging,scherer2009dynamic}, seeing the set of all possible emotional states as an abstract n-dimensional space, in which each point $\vec{x}(t) \in \mathbb{R}^n$ corresponds to a specific state. Its evolution in a continuous-time system is dictated by the following system of differential equations: 
\begin{equation}
\frac{d\vec{x}}{dt} = \vec{F}(\vec{x},\vec{u},\theta)
\end{equation}
where external inputs and individual-difference parameters (represented above by $\vec{u}$ and $\theta$) determine the evolution of the modeled state. In the present context, the equations should be interpreted as explicit hypotheses about one possible feedback pathway within cross-neurotype interaction, rather than as a complete representation of empathy or as an empirically established account of autistic and non-autistic behavior. 

We first consider a dyadic interaction between two agents, labeled Agent A and Agent NT to reflect the intended autistic-non-autistic application. These labels are convenient for simulations but should not be interpreted as homogeneous diagnostic prototypes \citep{apa2013dsm5,who2019icd11}. The psychologically relevant quantities are the agents' communication profiles, expectations, and responses, together with the compatibility between them. The informational exchanges between these agents are binarily classified as either ``verbal" or ``nonverbal". This simplistic communicational taxonomy is surprisingly useful, as it allows for explicit modeling of channel preference for signal output.  

\begin{figure}[H]
\centering
\includegraphics[width=0.9\textwidth]{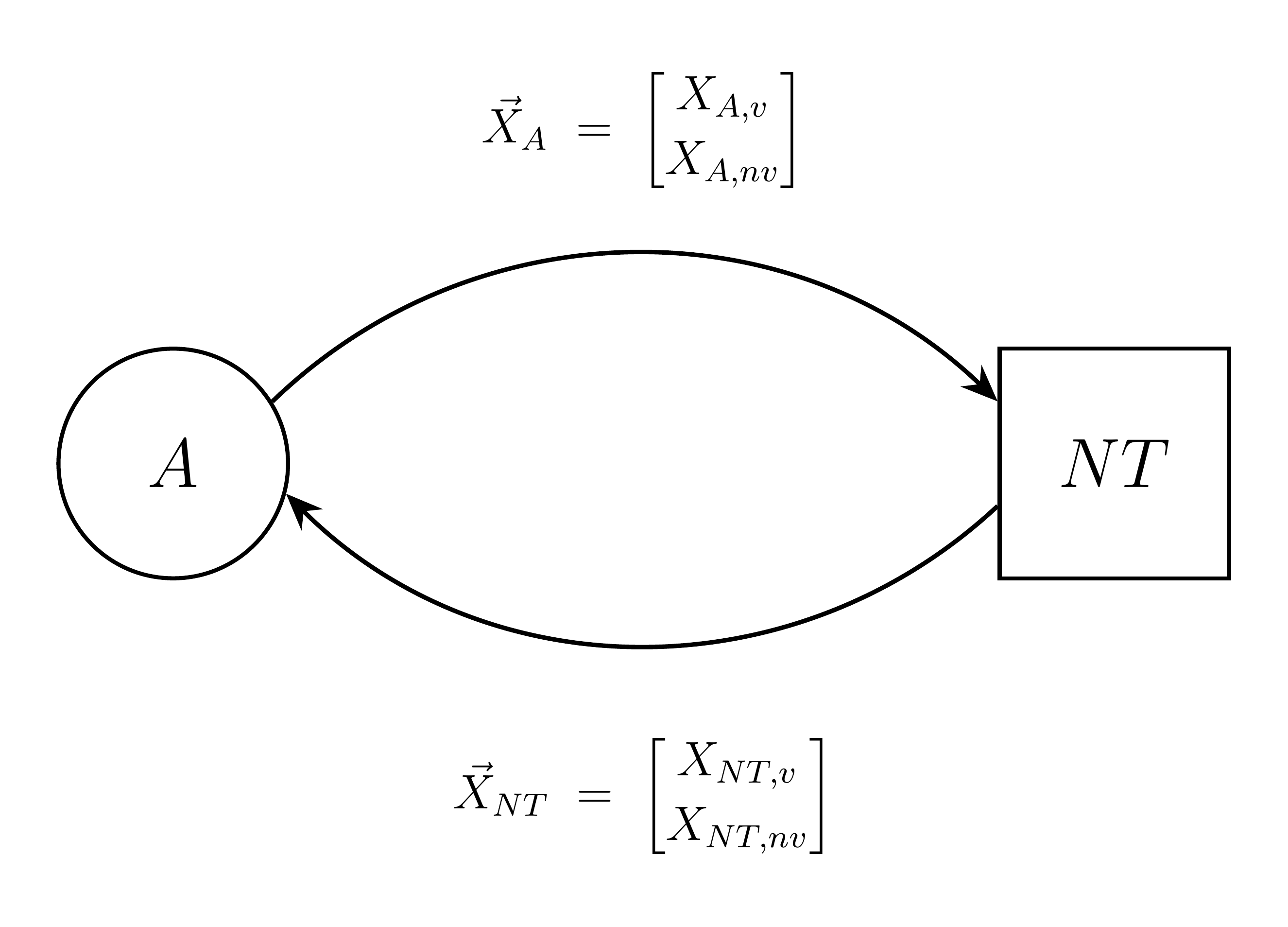}
\caption{A schematic diagram of the exchanges of verbal and nonverbal empathy outputs between Agent A and Agent NT.}\label{fig1}
\end{figure}

To appropriately study this system, we define an abstract two-dimensional space of empathy output, defined by the dimensions `Verbal Empathy Output’ and `Nonverbal Empathy Output’. We seek to analyze the trajectory of the empathy signals emitted by each agent as they evolve in this space. Every point in this space can therefore be described as a two-component empathy output vector: 

\begin{equation}
{\vec{X}}_{i} = \begin{bmatrix} X_{i,v}  \\
 X_{i,nv} \end{bmatrix} \in [ 0,100 ] \times [ 0,100 ]
\end{equation}
where $i \in \{A,NT\}$, $X_{i,v}$ denotes the empathy emitted by Agent $i$ to the other Agent through the verbal channel, and $X_{i,nv}$ denotes the empathy emitted by Agent $i$ through the nonverbal channel. The magnitude of each of these two signals is mapped to a [0-100] numerical scale. Figure \ref{fig1} presents a schematic of this interaction. 

The current model contains three functions for each agent: an empathy-perception function, a defensive-state function, and an empathy-output function. These functions specify a feedback pathway in which perceived shortfalls increase defensivity and defensivity reduces subsequent output. In the broader framework of the communication-cognition interface, this proposed pathway should be considered alongside interaction cost, affective resonance, motivation, and reciprocal accommodation, which remain to be incorporated into the mathematical architecture. Nevertheless, the reductionist approach taken here is valuable for it allows us to illustrate in analytical and numerical approaches the relevance of channel bias in social communication.  

The empathy perception function $p_i(X_{j,v};X_{j,nv};\phi_i)$ takes the verbal and nonverbal empathy signals each individual receives from the other agent ($i\neq j$), as well as the agent’s specific parameters $\phi_i$ indicating how verbal and nonverbal channel signals are weighted during the interaction (i.e., whether one channel is preferred to the other, if there are cross-channel interactions or synergies, or if there are lower or upper thresholds or limits to perception) and combines them into a single scalar number, defined as the registered empathy ($RE_i$): 

\begin{equation}
RE_{i} = p_{i}(X_{j,v};X_{j,nv};\phi_{i}),\qquad(i \neq j)
\end{equation}
The perceived input is compared with an expected level of empathy represented on the same scale. Expectations may depend on context, relationship history, communication norms, prior cross-neurotype experience, and the observer's learned model of how empathy is usually expressed. Here we introduce the empathy gap registered by each agent: 

\begin{equation}
\Delta_{i} = EE_{i} - RE_{i}
\end{equation}
The existence of a nonzero empathy gap, therefore, represents a discrepancy between received empathy signals and the observer's expectation, though not necessarily an absence of empathic intention in the partner. In this context, positive gaps reflect shortcomings with respect to the expected intensity of empathy by Agent $i$, while negative gaps indicate Agent $i$ is registering more empathy than expected. The model then introduces an internal state termed `defensivity'. Operationally, this variable captures the assumption that a perceived empathy shortfall can increase guardedness, antagonism, fear, or withdrawal. While these states should not be treated as psychologically equivalent, they all share an intrinsic negative valence within the context of human cognitive-emotional processing. Therefore, defensivity is retained here as a specific modeled pathway to serve as a proof-of-principle that empathic bonding breakdown (or ``empathy collapse") can occur due to verbal/nonverbal channel preferences. However, we explicitly do not defend that this is the only mechanism leading to reduced empathy output or reduced engagement, as it can occur through cognitive fatigue, high interaction cost, low affective reward, or declining motivation.  This defensivity state evolves in discrete steps $n\to n+1$, according to the rule 

\begin{equation}
D_{i}[ n + 1 ] = D_{i}[ n ] + y_{i}(\Delta_{i}[ n + 1 ];\psi_{i}) - \lambda_{i}D_{i}[ n ]
\end{equation}
where $n=0,1,2,\dots$ serves as an index for each step. As we can see, $y_i(\Delta_i[n+1];\psi_i)$ represents the increment function that maps the current gap measured immediately before to an increase in the defensivity. $\psi_i$ indicates the agent-specific parameters for the $y$ function, and $\lambda_i\in [0,1]$ represents a decay or damping parameter to incorporate forgetting effects of this internal state \citep{thompson1994emotion,gross1998emerging}. 

At last, the empathy output functions $z_{i,v}(D_i)$ and $z_{i,nv}(D_i)$ map the agent’s defensivity value to new empathy outputs by the individual in question in the verbal and nonverbal channels, respectively, and the process starts again, now for the other individual in the dyad. 

As stated in the previous section, while autistic and non-autistic individuals show substantial within-group variability in cognitive and behavioral traits, there is widespread support in the literature that autistic communication places great emphasis on verbal processing and literality when compared to neurotypical styles of communication \citep{happe1993communicative,tagerflusberg2005language,vicente2023literal}. Though not considered a defining characteristic of autism, we expect this clinical observation to be empirically reproducible at group-level under direct measurement and take this notion as the underlying assumption of this model. This would be represented analytically in this model as Agent A weighting more the empathy emitted through the verbal channel by other individuals than the one emitted through the nonverbal channel, whereas a neurotypical would weight what they register from both channels roughly equally, or even give greater emphasis to the nonverbal channel. Therein lies the mechanism of our model for explaining the double empathy phenomenon: this differential processing of verbal and nonverbal information might make one individual seem less empathetic than their true intention, resulting in an antagonistic response from the other agent (a rise in the other agent’s defensivity, to employ the vocabulary defined here). Assuming the interaction does not cease immediately, this increase in defensivity would lead to a decrease in empathy output (both verbal and nonverbal), thereby triggering a feedback loop in which every agent lowers the empathy they express progressively, until their assessment of the other individual is almost completely hostile. We refer to this state as ‘empathy collapse.’ A pictorial representation of this scenario can be seen in Figure \ref{fig2} below. 

\begin{figure}[H]
\centering
\includegraphics[width=0.9\textwidth]{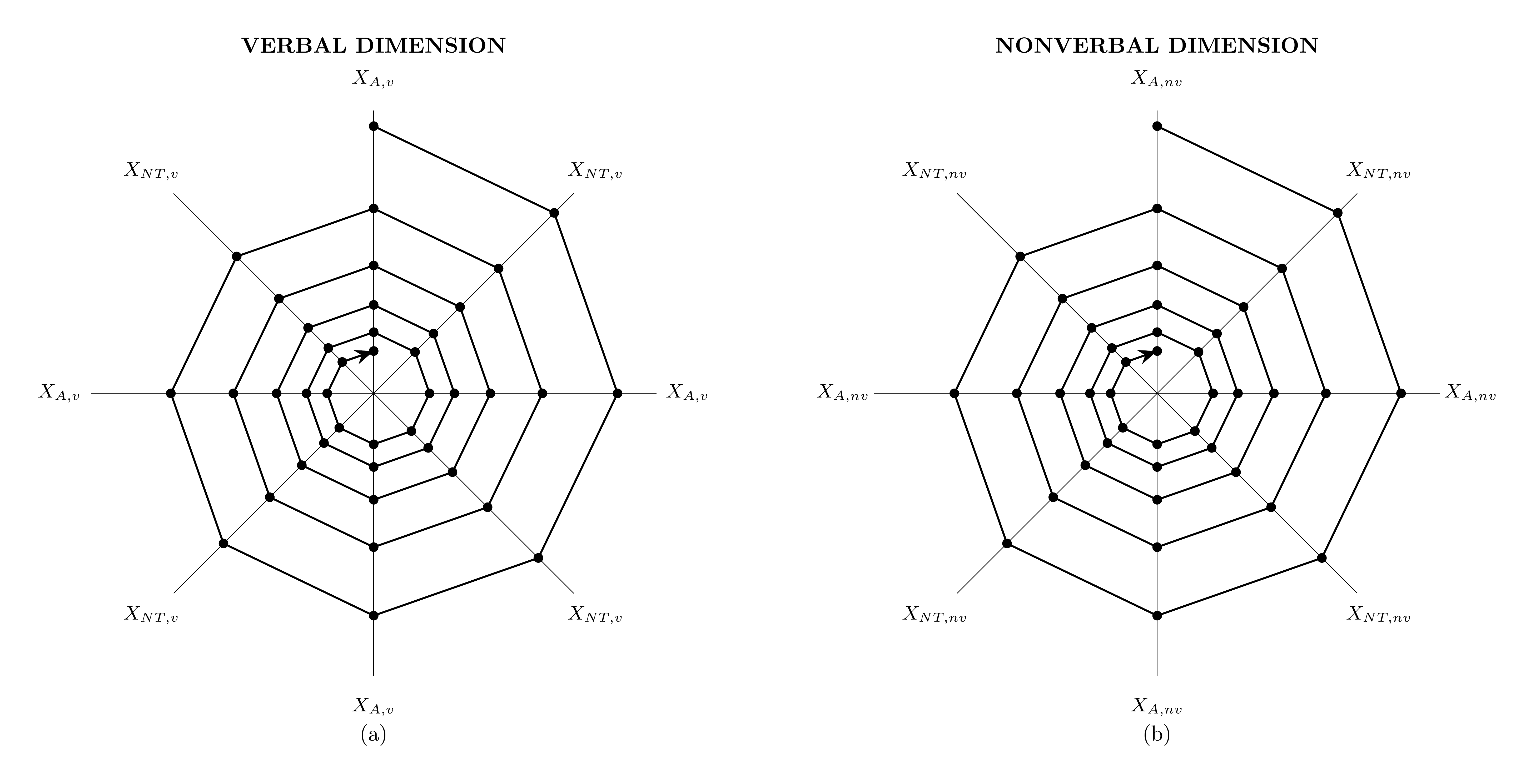}
\caption{A sketch of the evolution of the empathy outputs in the verbal dimension, in (a), and in the nonverbal dimension, in (b), for Agent A and Agent NT. The magnitude of the output for each agent decreases after each step, displaying this spiral dynamic. }\label{fig2}
\end{figure}

Interestingly, the dynamics of empathy collapse as plotted here resemble those of the famous Kindleberger spiral, named after economic historian Charles Kindleberger, for the collapse of world trade during the Great Depression \citep{kindleberger1973depression}. 

It is important to note that the mechanism examined here is therefore not autism itself, but a mismatch between how one partner expresses empathic information and how the other partner weighs the available channels when interpreting that information. Under particular parameter configurations, such production-perception incompatibility may cause empathic intention to be registered as weaker than intended, generating a perceived empathy gap and activating the modeled feedback pathway. In this sense, the framework is fundamentally diagnosis-agnostic: neurotype becomes relevant insofar as group-level communication tendencies alter the compatibility of particular profiles, rather than determining the interactional outcome. In addition, should the clinical observation of differential channel preference by neurotype not translate empirically to group-level differences in verbal and nonverbal weighting, the central assumption of the model would be challenged, a point to which we return below; this duality, therefore, constitutes a central falsifiable prediction of the model. 

An important aspect of the model is whether that particular combination of neuropsychological functions and numerical variables will, in fact, lead to empathy collapse or will stabilize around some fixed point in the empathy space that still allows for effective communication and emotional bonding. The precise analytic format of the functions is not known a priori, and their shape and parameters may vary from person to person and/or social context. As such, it would be extremely useful to determine if the overall trajectory of empathy outputs is stable (in which case small perturbations shrink over time) or unstable (in which case these small perturbations grow after each step). In this way, local information for the flows of empathy outputs in this abstract space may inform us regarding the global dynamical behavior of the social interaction. 

The only actual dynamical state for each of the agents in the model is their defensivity; all other quantities described above refer to instantaneous outputs based on said defensivities and generated by functions. Therefore, we may characterize the system by the defensivity state 

\begin{equation}
\vec{D} = \begin{bmatrix} D_{A}  \\
 D_{NT} \end{bmatrix}
\end{equation}

Formally, a fixed point in a dynamical system is a point that is mapped to itself by an evolution function. It is said stable if a small perturbation to the solution that led to the fixed point in the first place would shrink over time, thereby attracting orbits close to it. Otherwise, it is said unstable \citep{strogatz2015nonlinear}. We denote all functions evaluated at a fixed point 

\begin{equation}
{\vec{D}}^{*} = \begin{bmatrix} D_{A}^{*}  \\
 D_{NT}^{*} \end{bmatrix}
\end{equation}
(if it exists at all) by a star (e.g., $X_{A,v}^*$, $RE_{NT}^*$, $\Delta_A^*$ etc). In order to assess the stability of a given fixed point in the social interaction described by this model (that is, other equilibrium solutions different from the empathy collapse scenario), we linearize the equations of evolution of the system for a particular point in the two-dimensional defensivity space. As shown above, the defensivity of each agent evolves as 

\begin{equation}
\begin{aligned} D_{A}[ n + 1 ] & = D_{A}[ n ](1 - \lambda_{A}) + y_{A}(\Delta_{A}[ n + 1 ]) = f_{A}(D_{A}[ n ])  \\
 D_{NT}[ n + 1 ] & = D_{NT}[ n ](1 - \lambda_{NT}) + y_{NT}(\Delta_{NT}[ n + 1 ]) = f_{NT}(D_{NT}[ n ]) \end{aligned}
\end{equation}

The Jacobian matrix $J$ of a vector-valued function assembles all of its first-order partial derivatives, generalizing the concept of derivative to higher dimensions. For the vector map 

\begin{equation}
\vec{f} = \begin{bmatrix} f_{A}  \\
 f_{NT} \end{bmatrix}
\end{equation}
which generates the evolution in discrete steps of the system’s defensivity, we have the following Jacobian: 

\begin{equation}
J = \begin{pmatrix} \frac{\partial f_{A}}{\partial D_{A}} & \frac{\partial f_{A}}{\partial D_{NT}}  \\
 \frac{\partial f_{NT}}{\partial D_{A}} & \frac{\partial f_{NT}}{\partial D_{NT}} \end{pmatrix}
\end{equation} 
As the $y$ functions do not depend on the current defensivity for either agent, the main diagonal of this matrix contains the terms $1-\lambda_A$ and $1-\lambda_{NT}$. For the off-diagonal terms, we have: 

\begin{equation}
\begin{aligned} J_{12} = \frac{\partial f_{A}}{\partial D_{NT}} & = \frac{\partial y_{A}(EE_{A} - p_{A}(X_{NT,v},X_{NT,nv},\phi_{A}))}{\partial D_{NT}} = - \frac{dy_{A}}{d\Delta_{A}}\frac{\partial p_{A}}{\partial D_{NT}}  \\
 J_{21} = \frac{\partial f_{NT}}{\partial D_{A}} & = \frac{\partial y_{NT}(EE_{NT} - p_{NT}(X_{A,v},X_{A,nv},\phi_{NT}))}{\partial D_{A}} = - \frac{dy_{NT}}{d\Delta_{NT}}\frac{\partial p_{NT}}{\partial D_{A}} \end{aligned}
\end{equation}
By defining the perceptual sensitivities of each agent to the other agent’s defensivity as 

\begin{equation}
\begin{aligned} S_{A \leftarrow NT} & = - \frac{\partial p_{A}}{\partial D_{NT}}  \\
 S_{NT \leftarrow A} & = - \frac{\partial p_{NT}}{\partial D_{A}} \end{aligned}
\end{equation}
we can write these partial derivatives presented above as 

\begin{equation}
\begin{aligned} \frac{\partial f_{A}}{\partial D_{NT}} & = y_{A}'S_{A \leftarrow NT}  \\
 \frac{\partial f_{NT}}{\partial D_{A}} & = y_{NT}'S_{NT \leftarrow A} \end{aligned}
\end{equation} 
as well as define a scalar index which measures the propensity of the system for instability, the loop-gain product: 

\begin{equation}
\mathcal{L} = y_{A}'\,S_{A \leftarrow NT}\; y_{NT}'\,S_{NT \leftarrow A}
\end{equation}

For a linear system in discrete time, local stability of a fixed point is determined by the Jury conditions \citep{jury1964ztransform}, which stipulate stability if both eigenvalues $\lambda_{1,2}$ (not to be confused with the damping factors of each agent’s defensivity) of the Jacobian matrix evaluated at that point lie within the unit circle on the plane. This is equivalent to the following conditions: 

\begin{equation}
\begin{aligned} & 1 + \operatorname{tr}(J^{*}) + \det(J^{*}) > 0  \\
 & 1 - \operatorname{tr}(J^{*}) + \det(J^{*}) > 0  \\
 & 1 - \det(J^{*}) > 0 \end{aligned}
\end{equation}
where $\operatorname{tr}(J^{*})$ and $\det(J^{*})$ denote the trace and the determinant of the Jacobian matrix computed on that particular fixed point, respectively. In particular, using the notation defined above, the determinant of the Jacobian can be written as 

\begin{equation}
\det(J) = (1 - \lambda_{A})(1 - \lambda_{NT}) - \mathcal{L}
\end{equation}

Having outlined the general structure of the model we propose to explain the verbal-nonverbal dimension of the double empathy problem, we move on to a question of existence: are there specific functions $p_i$, $y_i$, and $z_i$ that, using appropriate parameters, result in the collapse of empathy output in a dyadic social exchange? Neither the analytical form of such maps nor the actual values of such parameters are known a priori. However, reasonable suppositions and previous knowledge do act to constraint this function hyperspace. For example, as stated above, it is widely known that autistic individuals weight verbal information much more heavily than nonverbal information, allowing us to conclude that the `weights’ of the verbal input for Agent A’s $p$ function must be numerically superior to the weights for nonverbal input, whereas Agent NT’s weights are more equilibrated. Furthermore, human social dynamics and the internal states derived thereof (such as defensivity) are unlikely to be adequately modeled by linear functions with sharp, discontinuous transitions. 

To further simplify the search for adequate maps, we can impose an ad hoc stipulation that defensivity has a sigmoidal evolution in a hostile interaction, which nevertheless seems probable: a sigmoid has a small but increasing slope. At some point, however, the slope’s rate of change goes to zero and then negative, leading to the curve’s overall saturation. This seems appropriate to model how one’s wariness increases when facing an antagonistic partner. There is an initial shock to the lack of empathy registration, but the defensivity does not immediately spike, as it takes some rounds of interaction for someone to compose a theory of mind attributing hostility to another person. Eventually, a person reaches such an understanding and the increase of defensivity is very rapid, until it reaches a final saturation, when the social fabric of the interaction has ruptured. 

We propose an implementation of the model with the $p_i$ function as a weighted linear combination: 

\begin{equation}
RE_{i} = p_{i}(X_{j,v};X_{j,nv};c_{i,v};c_{i,nv}) = c_{i,v}X_{j,v} + c_{i,nv}X_{j,nv}
\end{equation}
where $c_{i,v}$ ($c_{i,nv}$)represents the weight given by Agent $i$ to the signal they register through the verbal (nonverbal) channel. In this simple linear formulation, the total perceived empathy is an additive combination of the two channels, with no interaction or synergy effects. For this version of the model, the empathy one Agent can send to the other Agent is any vector belonging to the Cartesian product $[0,100]\times[0,100]$, and is mapped by the perception functions to a scalar $RE_i \in [0,100]$. Therefore, it is necessary that $c_{i,v} + c_{i,nv}=1$. To reduce the number of parameters, we set $c_{i,v}=c_{i,p}$, which of course means that $c_{i,nv}=1-c_{i,p}$.              

For the defensivity increment function $y_i(\Delta_i,D_i)$, we implement a linear function with saturation: 

\begin{equation}
y_{i}(\Delta_{i},D_{i}) = c_{y,i}(\Delta_{i} - \theta_{i})(1 - D_{i}/D_{sat,i})
\end{equation}
where the coefficient $c_{y,i}$ controls the sensitivity of defensivity to empathy gaps, with higher values creating stronger coupling between perceived shortfalls and defensive responses (this parameter can be interpreted as individual differences in emotional reactivity or sensitivity to social rejection); $\theta_i$ represents a lower threshold (specific to Agent $i$) for an empathy gap to increase their defensivity (which is interpreted as the individual ability to suppress hostile responses to minor rejection); and $D_{\text{sat},i}$ represents Agent $i$’s saturated value of defensivity, preventing it from growing unboundedly and which we believe adequately reflects human responses. As defensivity approaches this saturation threshold, the effective gain at each step is progressively reduced. From a modeling perspective, this term introduces state-dependent gain, creating a nonlinearity that both complicates and enriches the dynamics observed. 

For the form of the $z_i$ function, we implement the following function: 

\begin{equation}
\begin{aligned} X_{i,v} & = X_{\max,i,v} - c_{z,i,v}D_{i}  \\
 X_{i,nv} & = X_{\max,i,nv} - c_{z,i,nv}D_{i} \end{aligned}
\end{equation} 
where $X_{\text{max},i,v}$ and $X_{\text{max},i,nv}$ are the maximum empathy outputs for the verbal and nonverbal channels of Agent $i$, respectively, and $c_{z,i,v}$ and $c_{z,i,nv}$ are the output coefficients controlling the rate of empathy suppression for the verbal and nonverbal channels for Agent i per unit of defensivity. This linear relationship captures the intuition that defensive individuals express less empathy: as one becomes more guarded or hostile toward an interlocutor, one naturally reduces both verbal and nonverbal signals of care and connection. These outputs are obviously limited above by $X_{\text{max},i,v}$ and  $X_{\text{max},i,nv}$, and we proceed to cap them below at zero, as a negative amount of empathy lacks interpretation other than explicit hostility, which is not the concern of our model. 

The output function closes the feedback loop, with higher defensivity leading to lower empathy output which, when perceived by the other agent, creates a larger empathy gap for the recipient of the signal, driving further defensivity growth. This positive feedback structure, along with the perception mismatch caused by how individuals with different neurotypes weigh different channels, is the engine of empathy collapse. 

Below we describe the sequential loop of the output and perceptual dynamics of the model: 

STEP 1 - Agent A emits their empathy output: 

\begin{equation}
\begin{aligned} X_{A,v}[ 0 ] & = X_{\max,A,v} - c_{z,A,v}D_{A}[ 0 ]  \\
 X_{A,nv}[ 0 ] & = X_{\max,A,nv} - c_{z,A,nv}D_{A}[ 0 ] \end{aligned}
\end{equation}

STEP 2 - Agent NT registers these signals and computes their empathy gap: 

\begin{equation}
\begin{aligned} RE_{NT}[ 1 ] & = c_{p,NT}X_{A,v}[ 0 ] + (1 - c_{p,NT})X_{A,nv}[ 0 ]  \\
 \Delta_{NT}[ 1 ] & = EE_{NT} - RE_{NT}[ 1 ] \end{aligned}
\end{equation}

STEP 3 - Agent NT updates their defensivity: 

\begin{equation}
D_{NT}[ 1 ] = D_{NT}[ 0 ](1 - \lambda_{NT}) + c_{y,NT}(\Delta_{NT}[ 1 ] - \theta_{NT})(1 - (D_{NT}[ 0 ])/D_{sat,NT})
\end{equation} 

STEP 4 - Agent NT emits their empathy output: 

\begin{equation}
\begin{aligned} X_{NT,v}[ 1 ] & = X_{\max,NT,v} - c_{z,NT,v}D_{NT}[ 1 ]  \\
 X_{NT,nv}[ 1 ] & = X_{\max,NT,nv} - c_{z,NT,nv}D_{NT}[ 1 ] \end{aligned}
\end{equation}

STEP 5 - Agent A registers these signals and computes their own empathy gap: 

\begin{equation}
\begin{aligned} RE_{A}[ 1 ] & = c_{p,A}X_{NT,v}[ 1 ] + (1 - c_{p,A})X_{NT,nv}[ 1 ]  \\
 \Delta_{A}[ 1 ] & = EE_{A} - RE_{A}[ 1 ] \end{aligned}
\end{equation} 

STEP 6 - Agent A updates their defensivity: 

\begin{equation}
D_{A}[ 1 ] = D_{A}[ 0 ](1 - \lambda_{A}) + c_{y,A}(\Delta_{A}[ 1 ] - \theta_{A})(1 - (D_{A}[ 0 ])/D_{sat,A})
\end{equation}
This completes the first cycle of the interaction, $n=1$. The process repeats itself identically for the second cycle, $n=2$, and so on. 

We define another measure of the decay dynamics of the system, which we name the decay preference (or output asymmetry parameter): 

\begin{equation}
\rho_{i} = \frac{c_{z,i,v}}{c_{z,i,nv}}
\end{equation}
It is defined as the ratio of the decay coefficient for the verbal empathy output and the decay coefficient for the nonverbal empathy output. When $\rho_i<1$, verbal output is preserved relative to nonverbal output. When $\rho_i>1$, verbal output decays faster. 

To more efficiently explore the empathy and defensivity trajectories, and the overall phase portrait of the system, we perform numerical simulations of these interaction dynamics, using for the parameters values that conform to the suppositions stated at the beginning of this section. We adopt the following parameters: $\lambda_A = \lambda_{NT}=0.02$ for the natural decay rates, $c_{y,A}=c_{y,NT}=0.10$ for the defensivity gain coefficients, $EE_A=EE_{NT}=96$ for the expected empathies, $\theta_A = \theta_{NT} = 1$ for the lower thresholds for defensivity increase, and $D_{\text{sat},A}=D_{\text{sat},NT}=100$ for the saturation ceilings. We cap the term $(\Delta_i - \theta_i)$, used in the computation of the defensivity increase, to a maximum of 98 for both agents, to ensure the $y$ function is piecewise linear. To reflect the differential weighting of verbal vs nonverbal channels by Agent A and Agent NT, we use verbal perception weights $c_{p,A}=0.75$ (and therefore $1-c_{p,A}=0.25$) and $c_{p,NT}=1-c_{p,NT}=0.5$. We use $c_{z,A,nv}=c_{z,NT,nv}=1.2$. We simulate a total of 200 steps. In this formulation of the model and with these parameters, the registered empathies are given by 

\begin{equation}
\begin{aligned} RE_{A} & = 0.75 \times X_{NT,\text{verbal}} + 0.25 \times X_{NT,\text{nonverbal}}  \\
 RE_{NT} & = 0.50 \times X_{A,\text{verbal}} + 0.50 \times X_{A,\text{nonverbal}} \end{aligned}
\end{equation} 
Agent A weights verbal signals three times as highly as nonverbal signals; Agent NT has balanced perception. Each agent’s empathy output decays with defensivity, but channels may decay at different rates controlled by the parameter $\rho$: 

\begin{equation}
\begin{aligned} X_{i,\text{verbal}} & = 100 - \rho_{i} \times 1.2 \times D_{i}  \\
 X_{i,\text{nonverbal}} & = 100 - 1.2 \times D_{i} \end{aligned}
\end{equation}
Figure \ref{fig3} below presents a schematic of the model. 

\begin{figure}[H]
\centering
\includegraphics[width=0.9\textwidth]{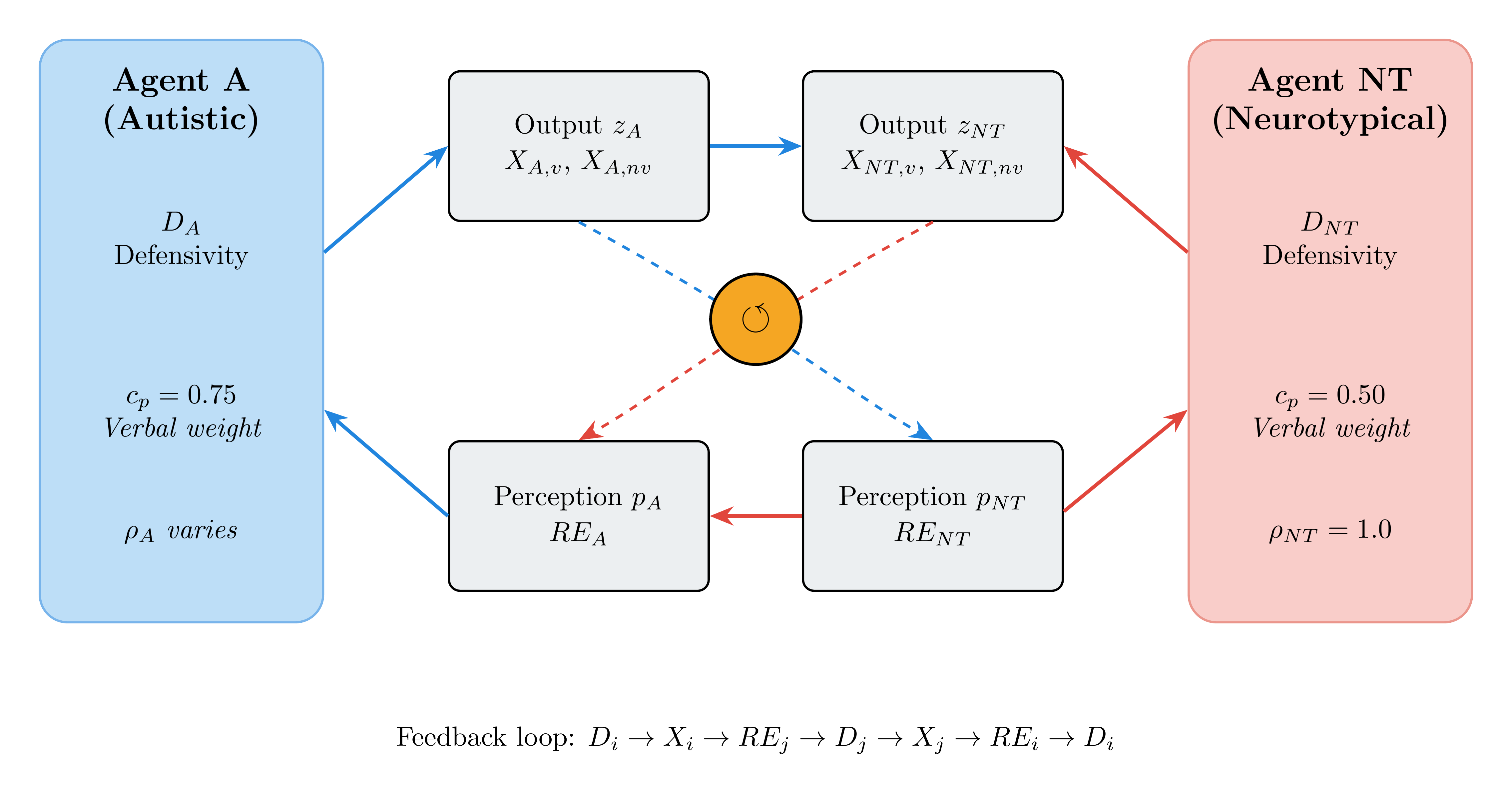}
\caption{A schematic representation of the version of the model used for the numerical simulations.}\label{fig3}
\end{figure}

We conducted 120 simulations systematically varying Agent A’s decay preference while holding Agent NT’s decay preference fixed at 1.0. This was done by setting $c_{z,A,nv}=c_{z,NT,v}=c_{z,NT,nv}=1.2$ while changing the value of $c_{z,A,v}$. We also systematically perturb Agent A’s initial defensivity while holding Agent NT’s initial defensivity fixed at 1.0. The variations adopted are: 

\begin{equation}
\begin{aligned} \rho_{A} & = \{ 0.5,\ 0.6,\ 0.7,\ 0.8,\ 0.9,\ 1.0,\ 1.1,\ 1.2,\ 1.3,\ 1.5,\ 1.75,\ 2.0\}  \\
 D_{A}[ 0 ] & = \{ 1.0,\ 2.0,\ 3.0,\ 4.0,\ 5.0,\ 5.5,\ 6.0,\ 6.5,\ 7.0,\ 8.0\} \end{aligned}
\end{equation}
This allows us to characterize the vulnerability of the modeled dyad to convergence toward a low-output state that we term `empathy collapse' following perturbation of Agent A's internal state. The term should not be interpreted as loss of empathic capacity, empathic concern, or emotional experience in either partner. Rather, it denotes a breakdown in the observable exchange and registration of empathic signals within the dyad. 

As the purpose of the current simulations is to propose a model whose mechanism could contribute to the double empathy problem, we restrict our analysis to an asymmetrical variation of parameters, which implies this does not yet constitute a complete test of bilateral interaction cost or reciprocal adaptation. The exclusive variation of Agent A's parameters should also not be understood as implying the instability originates preferentially from the autistic partner. A fuller bilateral analysis should vary homologous parameters in both agents and test whether equivalent incompatibilities generate comparable dynamics when the roles of the agents are reversed.

\section{Results}\label{sec:results}

Of 120 simulations, 20 resulted in collapse (16.7\% of the total). All collapses occurred in runs with $\rho_A \ge 1.0$. Table \ref{tab1} below presents the value of $\rho_A$ and $D_A[0]$ for which the collapses occurred. 

\begin{table}[h!]
\caption{Values of $\rho_A$ and $D_A[0]$ leading to empathy collapse across 120 simulations.}\label{tab1}
\begin{tabular*}{\textwidth}{@{\extracolsep\fill}ll@{\extracolsep\fill}}
\toprule
$\rho_A$ & Values of $D_A[0]$ leading to collapse \\
\midrule
0.50--0.90 & None \\
1.00 & 8.0 \\
1.10 & 8.0 \\
1.20 & 7.0, 8.0 \\
1.30 & 7.0, 8.0 \\
1.50 & 6.5, 7.0, 8.0 \\
1.75 & 5.5, 6.0, 6.5, 7.0, 8.0 \\
2.00 & 5.0, 5.5, 6.0, 6.5, 7.0, 8.0 \\
\botrule
\end{tabular*}
\end{table}

When Agent A preserves verbal output ($\rho_A\le 0.9$), no collapse occurred at any tested initial condition, including $D_A[0]=8.0$. The system remained unconditionally stable within the tested range, because when Agent A becomes defensive but has slow verbal expression degradation ($\rho_A<1$), Agent NT continues to perceive adequate empathy via the verbal channel. This prevents Agent NT from becoming defensive, breaking the positive feedback loop before it can escalate. The system possesses two qualitatively distinct fixed points. Between the two stable fixed points lies a separatrix that divides the state space into basins of attraction. Initial conditions below the separatrix converge to the low-defensivity equilibrium; those above it collapse to the high-defensivity state. The low-defensivity fixed point occurs when both agents’ perceived empathy exceeds the expected threshold ($RE_i\ge EE_i$), causing defensivity to decay toward zero. Setting the empathy gap to zero and solving for the steady state yields $D_A^* \approx D_{NT}^* \approx 0$. At this equilibrium, both agents express full empathy ($X_{\text{verbal}} = X_{\text{nonverbal}} = 100$), perceive full empathy ($RE_A = RE_{NT}=100$), and experience no empathy gap. The Jacobian matrix in this particular scenario is diagonal because the empathy gaps are zero, so there is no feedback coupling. The eigenvalues are both equal to $(1-\lambda)=0.98$, indicating asymptotic stability. Small perturbations decay exponentially with time constant $-\frac{1}{\text{ln}(1-\lambda)}\approx 50$ steps. However, this local stability does not guarantee global stability. The basin of attraction of the low-defensivity fixed point is bounded by a separatrix beyond which trajectories are captured by the high-defensivity attractor. The location of this separatrix depends critically on $\rho_A$.  The high-defensivity fixed point emerges when the positive feedback loop drives both agents toward the saturation ceiling. At this collapsed equilibrium, $D_A^*$ and $D_{NT}^*$ have values around 80, varying slightly according to the specific choice of $D_A[0]$ and $\rho_A$. Both agents’ empathy outputs are nearly zero, perceived empathy is minimal, and the large empathy gap is balanced by the saturation term ($1-D/D_{\text{sat}}$) to produce a stable steady state. While the actual values of defensivity, empathy perception and output for both agents are numerically similar at full collapse, the trajectory to collapse shows greater experiential asymmetry. Agent A’s higher verbal weight means they experience the verbal channel collapse more acutely during the transition phase. Figures \ref{fig4} and \ref{fig5} below present two representative cases of simulations without empathy collapse and with empathy collapse, respectively. 
\newpage

\begin{figure}[h!]
\centering
\includegraphics[width=0.9\textwidth]{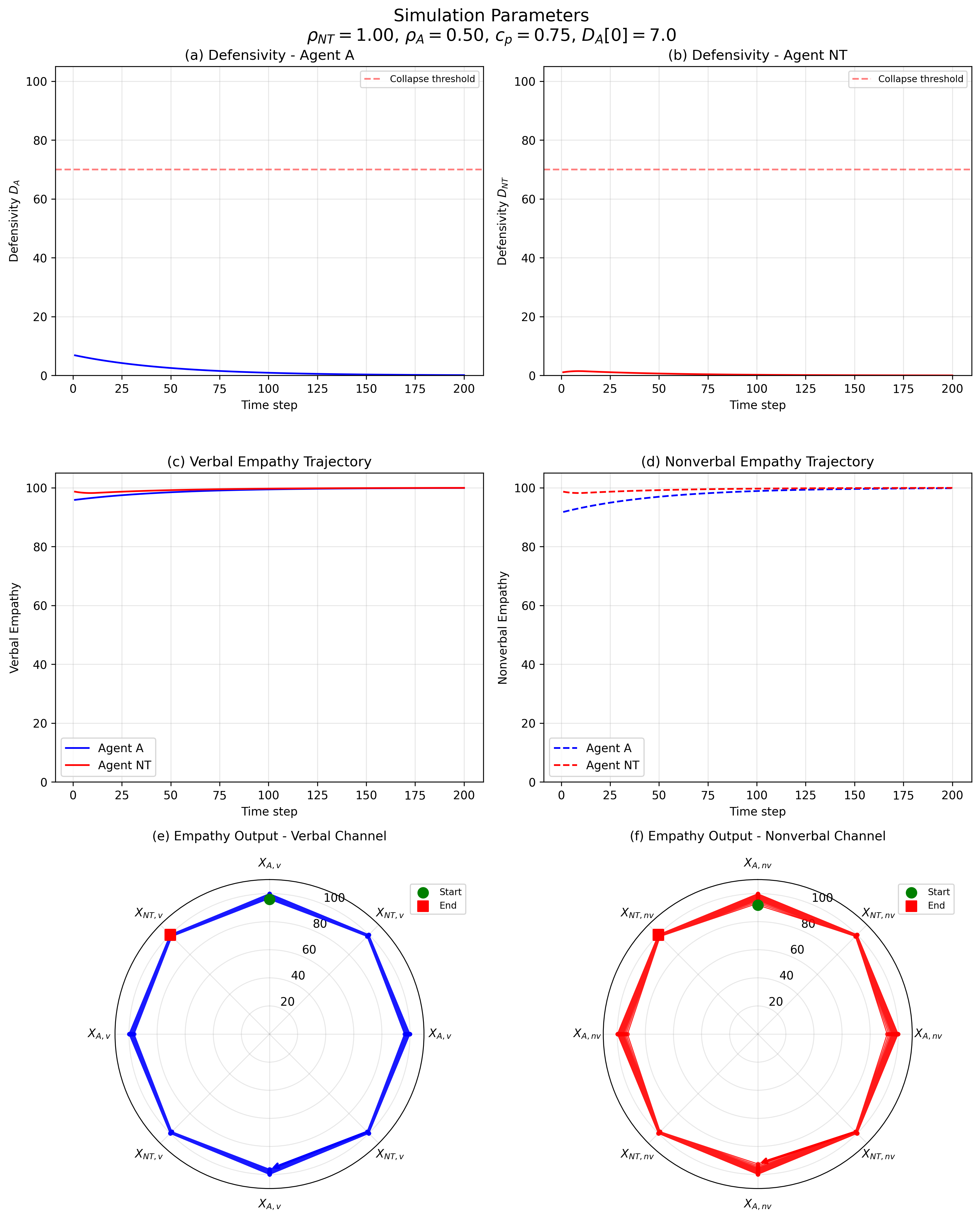}
\caption{Representative dynamics of defensivity and empathy output for a simulation with no collapse. (a,b) Defensivity of Agent A and Agent NT across 200 steps; the dashed line marks the collapse threshold. (c,d) Verbal and nonverbal empathy trajectories for both agents. (e,f) Empathy outputs in the eight-spoke representation of Figure 2, from start (green circle) to end (red square). Simulation parameters are given above the panels. }\label{fig4}
\end{figure}
\newpage

\begin{figure}[h!]
\centering
\includegraphics[width=0.9\textwidth]{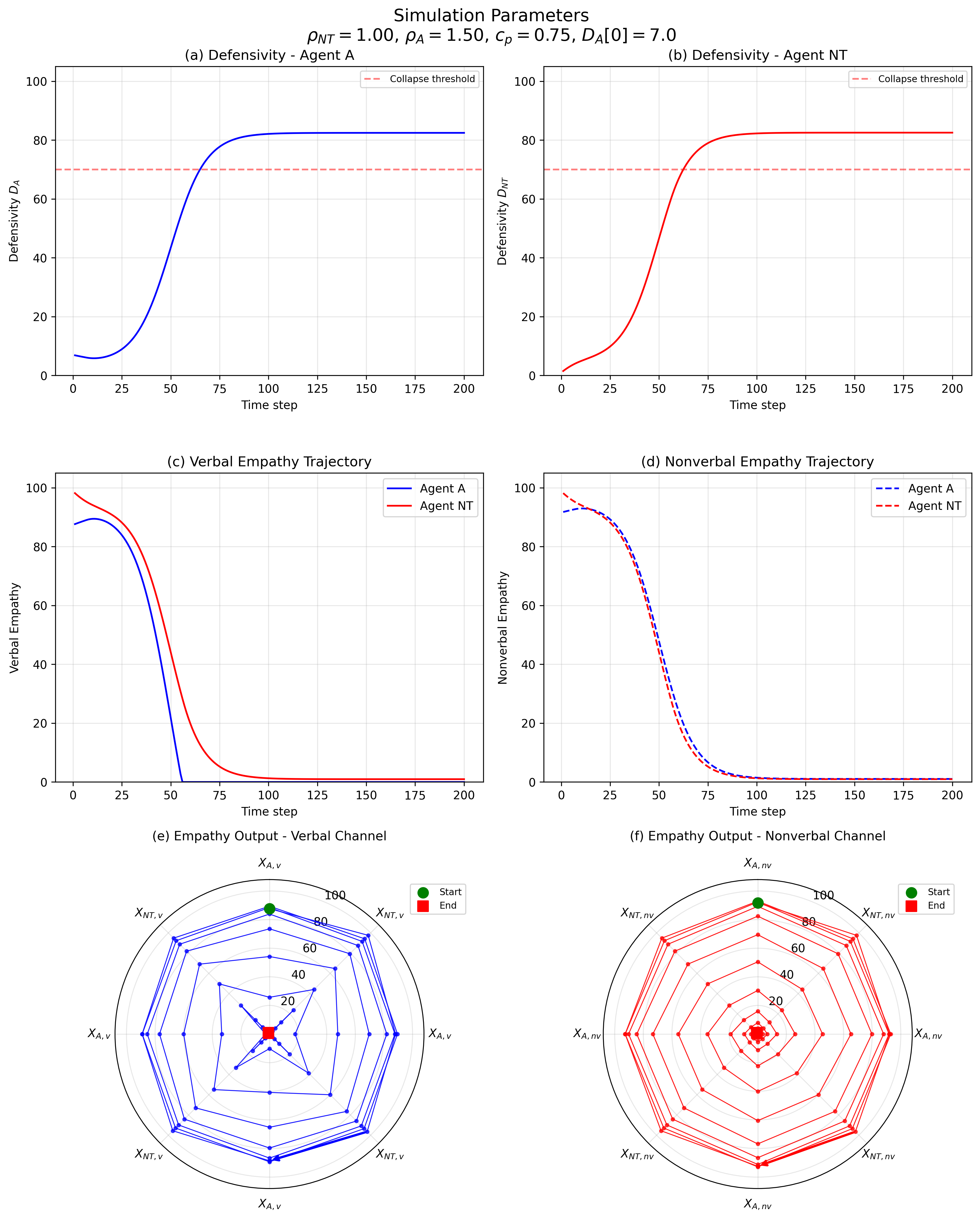}
\caption{Representative dynamics of defensivity and empathy output for a simulation with empathy collapse. (a,b) Defensivity of Agent A and Agent NT across 200 steps; the dashed line marks the collapse threshold. (c,d) Verbal and nonverbal empathy trajectories for both agents. (e,f) Empathy outputs in the eight-spoke representation of Figure 2, from start (green circle) to end (red square). Simulation parameters are given above the panels. }\label{fig5}
\end{figure}
\newpage

To understand why the output asymmetry parameter $\rho_A$ determines collapse vulnerability, we performed a systematic stability analysis of the simulations, as outlined by \citet{strogatz2015nonlinear}. This analysis reveals that $\rho_A$ affects system stability through two distinct mechanisms: it lowers the threshold at which the positive feedback loop activates, and it amplifies the strength of the feedback coupling between agents. 

The off-diagonal Jacobian elements capture the feedback coupling between agents. The element representing how Agent NT’s defensivity drives Agent A’s empathy perception is: 
\begin{equation}
\frac{\partial p_{A}}{\partial D_{NT}} = - (c_{p,A}c_{z,NT,v} + (1 - c_{p,A})c_{z,NT,nv}) = - ((0.75 \times 1.2) + (0.25 \times 1.2)) = - 1.2
\end{equation}
This sensitivity is independent of $\rho_A$. It depends only on Agent A’s perception weights and Agent NT’s output coefficients, all of which are held constant across all simulations. For Agent NT perceiving Agent A, with $c_{p,NT}=1-c_{p,NT}=0.50$, and A’s output coefficients $c_{z,A,v}=\rho_A\times 1.2$ and $c_{z,A,nv}=1.2$, the sensitivity becomes: 

\begin{equation}
\frac{\partial p_{NT}}{\partial D_{A}} = - ((0.50 \times \rho_{A} \times 1.2) + (0.50 \times 1.2)) = - 0.6 \times (\rho_{A} + 1)
\end{equation} 
This sensitivity increases linearly with $\rho_A$. When A’s verbal output of empathy decays faster than their nonverbal counterpart (higher $\rho_A$), Agent NT experiences a larger drop in perceived empathy for any given increase in Agent A’s defensivity. 

The loop gain $\mathcal{L}$ quantifies the total amplification around the feedback cycle: a perturbation in $D_A$ affects $RE_{NT}$, which drives $D_{NT}$, which affects $RE_A$, which returns to influence $D_A$. Mathematically, we have $\mathcal{L}=J_{12}\times J_{21}$. Substituting the expressions derived above, we find: 

\begin{equation}
\begin{split}
\mathcal{L} &= \left[ c_{y,A} \times 1.2 \times \left( 1 - \frac{D_{A}}{D_{sat,A}} \right) \right] \times \left[ c_{y,NT} \times 0.6 \times (\rho_{A} + 1) \times \left( 1 - \frac{D_{NT}}{D_{sat,NT}} \right) \right] \\
&= 0.0072 \times (\rho_{A} + 1) \times \left( 1 - \frac{D_{A}}{100} \right) \times \left( 1 - \frac{D_{NT}}{100} \right)
\end{split}
\end{equation}
This expression reveals the central finding: loop gain scales linearly with $(\rho_A+1)$. At equal defensivity levels ($D_A=D_{NT}=1$), the loop gain and the value of the greater eigenvalue are given by Table \ref{tab2} below.

\begin{table}[h!]
\caption{Loop gain $\mathcal{L}$ and maximum eigenvalue as functions of $\rho_A$ with equal defensivity ($D_A = D_{NT} = 1$).}\label{tab2}
\begin{tabular*}{\textwidth}{@{\extracolsep\fill}lcc@{\extracolsep\fill}}
\toprule
$\rho_A$ & $\mathcal{L}$ & Maximum eigenvalue \\
\midrule
0.50 & 0.01058508 & 1.08288382 \\
1.00 & 0.01411344 & 1.0988 \\
1.50 & 0.0176418 & 1.11282244 \\
2.00 & 0.02117016 & 1.12549969 \\
\botrule
\end{tabular*}
\end{table}

The positive feedback loop remains dormant when both empathy gaps are zero or negative (i.e., when $RE\ge EE$ for both agents). The feedback activates only when perceived empathy falls below the expected threshold. For Agent NT, the gap becomes positive when $EE_{NT}-RE_{NT}\ge 0$, which, for these parameters, is equivalent to: 

\begin{equation}
D_{A} > \frac{6.67}{(\rho_{A} + 1)}
\end{equation}
Therefore, higher $\rho_A$ lowers this threshold. At $\rho_A=0.5$, feedback activates at $D_A>4.45$; at $\rho_A=2.0$, it activates at $D_A>2.22$. This earlier activation compounds with the stronger gain to produce the observed collapse threshold dependence. Higher $\rho_A$ thus creates a dual vulnerability: the feedback loop activates at lower defensivity levels, and once active, operates with greater gain. 

Local stability requires that all eigenvalues of the Jacobian have magnitude less than unity. We computed eigenvalues along the trajectories of all 120 simulations. At the low-defensivity equilibrium ($D_A\approx D_{NT} \approx 0$), the system is unconditionally stable with eigenvalues $\lambda_1 = \lambda_2 = 0.98$, reflecting the natural decay rate. 

During the collapse process, eigenvalues transiently exceed unity. For the representative case of $\rho_A = 1.5$ with $D_A[0]=7.0$ shown in Figure \ref{fig5}, the maximum eigenvalue reaches $\lambda_{\text{max}}=1.105257755$ at step 8, precisely when the feedback loop first activates with $\mathcal{L}=0.0162289$. This local instability drives the explosive growth in defensivity. As the system becomes trapped in the basin of the high-defensivity equilibrium (for this case $D_A^* = 82.462$ and $D_{NT}^* = 82.529$), eigenvalues return below unity ($\lambda_{\text{max}}\approx 0.997$ at step 51), indicating that the collapsed state is locally stable and the system will converge to the fixed point. 

We determined the separatrix between the stable and collapse basins through binary search, identifying the minimum initial defensivity $D_{A}[0]$ that leads to collapse (defined as either $D_A$ or $D_{NT}$ being greater than 70) for each value of $\rho_A$.  The results are presented in Table \ref{tab3}. 

\begin{table}[h!]
\caption{Collapse threshold $D_A[0]$ and basin size change as functions of $\rho_A$.}\label{tab3}
\begin{tabular*}{\textwidth}{@{\extracolsep\fill}lcl@{\extracolsep\fill}}
\toprule
$\rho_A$ & Collapse threshold $D_A[0]$ & Basin variation vs.\ $\rho_A = 1.0$ \\
\midrule
0.50 & 8.1052 & $+35.45\%$ larger stable basin \\
0.70 & 7.0976 & $+18.61\%$ \\
1.00 & 5.9837 & (reference) \\
1.30 & 5.1720 & $-13.57\%$ \\
1.50 & 4.7436 & $-20.73\%$ \\
2.00 & 3.9296 & $-34.33\%$ smaller stable basin \\
\botrule
\end{tabular*}
\end{table}

\section{Discussion}\label{sec:discussion}

\subsection{Mechanisms of Collapse}

Higher $\rho_A$ shifts the separatrix downward, shrinking the basin of attraction of the healthy equilibrium and expanding the collapse basin. As the loop gain $\mathcal{L}$ is a function of $\rho_A+1$, the correlation between the size of the basin of attraction and $\mathcal{L}$ is negative. In fact, this relationship is approximately inverse: a doubling of the loop gain roughly halves the size of the stable basin. 

As such, the stability analysis we outlined in this section identifies two complementary mechanisms through which $\rho_A$ determines collapse vulnerability: 

First, $\rho_A$ controls the coupling strength from Agent A to Agent NT. When Agent A’s verbal output decays rapidly under stress (high $\rho_A$), any increase in Agent A’s defensivity produces a larger reduction in Agent NT’s perceived empathy. The coupling coefficient $J_{21}=0.06\times(\rho_A+1)\times(1-D_{NT}/100)$ increases by 50\% as $\rho_A$ goes from 1.0 to 2.0. 

Second, $\rho_A$ determines when the feedback loop engages. Higher $\rho_A$ means Agent NT’s empathy gap becomes positive at lower levels of Agent A’s defensivity, allowing the positive feedback cascade to initiate from less provocation. 

These mechanisms compound: high-$\rho_A$ individuals not only trigger the feedback loop more easily but also drive it more forcefully once triggered. This explains the dramatic protective effect of verbal preservation ($\rho_A<1$), as such individuals neither easily activate nor strongly drive the collapse cascade, resulting in stable interactions across nearly all tested initial conditions. This sharp boundary aligns with the loop gain analysis: below $\rho_A\approx 1.0$, the feedback coupling is weak enough that natural decay ($\lambda_A =\lambda_{NT}=0.02$) dominates, preventing runaway amplification regardless of initial perturbation magnitude. 

\subsection{Experimental Designs for Parameter Measurement}

Considering the current lack of empirical data regarding the time evolution of empathy output during social interactions for human subjects, and how this may be altered by consideration of the canonical communication profiles of distinct neurotypes, the model in its current iteration lacks experimental verifiability, rendering questions of numerical validity moot and making falsifiability a major concern. In addition to that, as stated above, this model focuses exclusively on the verbal-nonverbal dimensional of empathic bonding for dyads; thus, the observation of empathy collapse does not imply the underlying mechanism depended on verbal and nonverbal channels of communication. Nevertheless, we list several trial designs below allowing for the measurement of the pertinent parameters. Controlling for other processes of social and emotion friction within interactions should allow us to ascertain whether the dynamics of the verbal-nonverbal dimension of communicative mismatch and its contribution to the double empathy problem behave as the model predicts or not.  

\subsubsection{Design 1: $c_p$}
The perception weight $c_p$ represents the relative attention an individual gives to verbal versus nonverbal empathy signals 	\citep{baroncohen2004empathy}. The model assumes autistic individuals weight verbal signals more heavily while neurotypical individuals have more balanced perception. For the simulations carried out, we used $c_{p,A}=0.75$ and $c_{p,NT}=0.50$. 

\paragraph{Measurement Protocol}
Participants view video stimuli of actors expressing empathy with independently manipulated verbal content (supportive statements) and nonverbal content (facial expressions, tone, posture). Using a factorial design crossing high/low verbal empathy with high/low nonverbal empathy, participants rate perceived empathy on a continuous scale after each video. Regression analysis yields individual perception weights: 

\begin{equation}
RE_{\text{rated}} = c_{p}X_{\text{verbal}} + \left( 1 - c_{p} \right)X_{\text{nonverbal}} + \epsilon
\end{equation}
The coefficient $c_p$ is estimated for each participant. The model predicts that autistic participants will, on average, show higher $c_p$ values (verbal weighting) than neurotypical participants, despite inherent individual variability. A failure to observe this difference, or observation of the opposite pattern on the aggregate of the neurotypical group and the autistic group, would challenge a core model assumption. An addition to this protocol would be to employ eye-tracking during conversational videos, measuring attention allocation to face and body posture/movement (nonverbal) -versus speech/captions (verbal), and therefore providing a behavioral correlate of perception weights that does not rely on self-report and could be combined with the original data.

\subsubsection{Design 2: $\rho$}
Parameter $\rho$ measures whether verbal or nonverbal channels of empathy output degrade faster. When $\rho>1$, verbal output decays faster than nonverbal; when $\rho<1$, verbal expression degrades more slowly than nonverbal expression. 

\paragraph{Measurement protocol}
Participants engage in a structured interaction task with a confederate while undergoing a stress induction, such as time pressure, cognitive load, or mild social evaluative threat. Empathy expression is coded from video recordings at baseline and under stress conditions. Verbal empathy is operationalized as explicit supportive statements, acknowledgments, and verbal expressions of understanding. Nonverbal empathy is coded from facial expression warmth, postural orientation, and paralinguistic features such as tone and prosody. Both channels are rated on 0-100 scales by trained coders blind to participant diagnosis. The output asymmetry ratio is calculated as: 
\begin{equation}
\rho = \frac{\Delta X_{\text{verbal}}/X_{\text{verbal},\text{baseline}}}{\Delta X_{\text{nonverbal}}/X_{\text{nonverbal},\text{baseline}}}
\end{equation}
where $\Delta$ represents the change from baseline to stressed condition. Values greater than 1 indicate verbal-preferential decay; values less than 1 indicate nonverbal preferential decay. The model does not require that autistic individuals have any particular $\rho$ value -- rather, it predicts that an individual’s $\rho$ value will moderate their vulnerability to empathy collapse regardless of diagnostic status.

\subsubsection{Design 3: $D[0]$}
Initial defensivity represents the baseline level of guardedness or emotional reactivity an individual brings to an interaction. This parameter likely varies both between individuals and within individuals across contexts.

\paragraph{Measurement protocol}
Pre-interaction measures can include state anxiety scales, physiological arousal (heart rate variability, skin conductance), and situational factors (relationship history with partner, stakes of interaction). A composite defensivity index can be constructed and validated against early-interaction behavioral coding of defensive behaviors (interrupting, dismissing, topic avoidance). For experimental manipulation, initial defensivity can be induced through false feedback about the interaction partner (e.g., ``your partner rated your previous response as unhelpful") or through priming procedures. 

\subsection{Experimental Designs for Model Falsification}

Below, we enumerate several general proposals of experimental behavioral protocols that would allow researchers to falsify the quantifiable predictions of the model as it stands. 

\begin{criterion}[Perception weight distribution]
\textit{\newline Prediction.} Autistic individuals will show, on average, higher verbal perception weights ($c_p$) than neurotypical individuals, with distributions showing meaningful separation.

\noindent\textit{Falsification.} If empirical measurement reveals (a) no significant difference in $c_p$ between groups, with complete overlap in distributions and no central tendency difference, or (b) autistic individuals showing on average lower $c_p$ than neurotypical individuals, the perception asymmetry assumption would be falsified. 
\end{criterion}

\begin{criterion}[Collapse threshold modulation by $\rho$]
\textit{\newline Prediction.} Among individuals with high verbal perception weights ($c_p>0.6$), those with $\rho<1$ (those whose verbal empathy output decays slowly compared to nonverbal output) should show stable interactions (constrained to control for other forms of interaction cost or affective resonance) across a wide range of initial conditions, while those with $\rho>1$ (verbal-decaying output) should show collapse at lower thresholds.

\noindent\textit{Falsification.} If individuals with $\rho<1$ show equal or greater collapse rates than those with $\rho>1$, controlling for initial defensivity, the output asymmetry mechanism for this version of the model would be falsified. Similarly, if collapse rates are independent of $\rho$ entirely, this component of the model fails. 
\end{criterion}

\begin{criterion}[Asymmetric experience in collapse dyads]
\textit{\newline Prediction.} In dyads that reach the collapsed state, the partner with higher $c_p$ should report lower perceived empathy than the partner with lower $c_p$ when measured simultaneously.

\noindent\textit{Falsification.} If collapsed dyads show symmetric perceived empathy deficits regardless of perception weights, or if the low-$c_p$ partner consistently reports greater deficits, the differential experience mechanism would be falsified. 
\end{criterion}

\begin{criterion}[Protective effect of verbal maintenance]
\textit{\newline Prediction.} Training interventions that specifically increase verbal empathy expression under stress (reducing $\rho$ toward values below 1) should reduce collapse rates in subsequent interactions.

\noindent\textit{Falsification.} If such interventions successfully modify $\rho$ but show no effect on collapse rates, or if interventions targeting nonverbal expression are equally or more effective, the model’s emphasis on verbal channel preservation as a possible mitigator of collapse would be falsified. 
\end{criterion}

A definitive test of these four falsification hooks would employ a longitudinal dyadic design: 

(i) Phase 1 (Parameter Estimation): Measure $c_p$ and $\rho$ for each participant using the protocols described above. 

(ii) Phase 2 (Interaction Task): Pair participants in structured conflict-resolution or support-seeking tasks with manipulated initial defensivity (through false feedback or priming). Record full interactions with behavioral coding of empathy channels and continuous self-report of perceived partner empathy. 

(iii) Phase 3 (Outcome Assessment): Classify interactions as stable or collapsed based on behavioral and self-report criteria. Test whether collapse rates follow the predicted pattern based on measured parameters. 

We emphasize, once more, that the measurements, experimental designs, and falsification hooks outlined here intend to extract how the channel preference might impact the double empathy problem, not to give a full account of the dynamics of dyadic empathic bonding or how it is transformed by neurodivergence. As such, some experiments would deliberately not seek the replication of naturalistic interaction scenarios, but to isolate channel dynamics from the confounding effects of other processes, which will be analyzed in the following section. Only after the individual analysis of each one of these dimensions of dyadic bonding and its failures should one seek to understand their combined synergistic effects and how meaningful the channel-preference mismatch is to realistic cross-neurotype empathic signaling and bonding. 

\subsection{Model Limitations}

Within the selected parametrization, the loop-gain analysis identifies combinations of output decay profile and initial defensivity that dictate the convergence toward the model's low-output attractor. However, the model is limited to just one dimension of dyadic bonding -- the communicative mismatch of channel preferences --, focuses solely on the downward trajectory toward interaction rupture and its relevance to the double empathy problem is based upon applying this general framework to the archetypal communicative profiles of distinct neurotype, thereby not incorporating factors that explain individual heterogeneity.  

In addition, the parameters used are theoretically motivated but not empirically calibrated, and the unilateral variation of Agent A's parameters -- motivated by concerns regarding the extension of this work -- limit the breadth of insight gained. Below, we provide a detailed account of social processes that impact emotional dyadic bonding that were not incorporated in this model and therefore present timely opportunities for future work. 

The model proposed here analyzes the dynamics of empathic bonding through the lens of communicational channels preferences. In doing so, many other variables known to causally influence dyadic bonding are deliberately neglected, as the objective is not to perfectly replicate every possible parameter variation but to establish that channel preferences and communicative mismatch can replicate the double empathy problem as it is observed clinically. The incorporation of more variables would require the analysis of evolution equations and trajectories in high-dimensional spaces, which is computationally costly and analytically unviable, besides frequently inhibiting an interpretative account of the model's causal structure. In addition, given the variability observed in psychiatric populations, the model focuses on archetypical representations of the communication preferences of differing neurotypes, as any attempt at modeling necessitates foregoing some degree of data granularity. Nevertheless, a fuller account of dyadic bonding dynamics would first delineate simple models of dynamical evolution for each parameter, check their predictive powers against empirical empathy measurements, and then proceed with incorporation into a unified model. The task then becomes to weight whether this lower mathematical tractability and greater difficulty in mechanism interpretability is offset by more accurate replication of clinical observations. We outline below several parameters overlooked in the current version of the model  

Naturally, communicative mismatch exists beyond the verbal/nonverbal duality, and it is expected to increase interaction cost. For the autistic partner, cost may arise from indirect language, implicit expectations, rapid integration of conventional nonverbal cues, sensory demands, or the need to monitor and mask natural behavior. For the non-autistic partner, cost may arise when prosody, facial expression, timing, gaze, gesture, or conversational organization differs from familiar conventions. The costs are therefore bilateral but need not be equal or arise from the same source: each partner may expend effort for different reasons and to different degrees. Future formalizations should therefore represent interaction cost separately for both agents rather than treating it as a property of one neurotype.  

A second mechanism is affective resonance. Empathic behavior is partly oriented by rapid processes such as emotional contagion, embodied resonance, affective salience, and interpersonal synchrony. An unfamiliar expression may generate weaker or delayed resonance in a particular observer whose perceptual history has been calibrated to different expressive regularities, even when the communicator's internal emotion is intense. This is a compatibility problem between expression and learned perceptual expectations, not evidence that autistic expression is intrinsically less emotional. 

A third mechanism is social motivation. We define motivation as the willingness to persist, infer, clarify, repair, and adapt within a particular interaction. It should be modeled dynamically as a function of perceived relational value, affective connection, expected success, and the cognitive-emotional cost to sustain the interaction, and it must be conceptualized as a dynamic property of the interaction itself rather than as a fixed characteristic of either neurotype. High motivation may sustain engagement despite substantial mismatch; low motivation in either partner may produce withdrawal without heightened defensivity. The model as it stands focuses on emotional perception and bonding, not engagement, which could persist regardless of internal state. However, increased motivation would extend engagement, creating more possibilities for affective reconnection.  

These processes converge in dyadic social coupling, defined here as the evolving degree of mutual predictability, coordination, affective connection, and communicative efficiency achieved by a pair. Low coupling can create a self-reinforcing cycle of high cost, low resonance, reduced motivation and expected success, and declining accommodation. Conversely, reciprocal clarification, familiarity, successful repair, and adaptation may lower cost and strengthen coupling. Importantly, coupling does not require communicative similarity or normalization of autistic behavior; it refers to the emergence of a workable shared interaction system. Future models should analyze the predictive power of these processes in describing dyadic social interaction and the double empathy problem, and check if a unified mathematical framework incorporating many or all these mechanisms is necessary to successfully model empathic bonding and its evolution. The relative relevance of each process and variable in determining bonding outcomes is a priori unknown. 

The mechanisms outlined above define the architecture of a broader social-cognitive framework, in which channel compatibility, bilateral interaction cost, affective resonance, and social motivation form the underpinnings of a general model of dyadic social coupling; each dimension can contribute independently or in consort with the others to determine if dyadic coordination will stabilize, deteriorate, or recover over time. 

Finally, the model's relevance to modeling autism communication rests on the assumption that the differential weighting of verbal and nonverbal emotional signals can represent with reasonable accuracy the previously reported findings of communication patterns in autistic individuals. However, it outlines a more general model of channel preference that can be useful in understanding autistic interactions, but that is fundamentally diagnosis-agnostic, replicating specific dynamics with statistical rather than deterministic accuracy. As remarked above, communication profiles within neurotypical and neurodivergent populations exhibit remarkable heterogeneity, and such variability will never be perfectly captured by an analytical model and its simplifying assumptions.

\section{Conclusion}\label{sec:conclusions}

This study developed and analyzed a dynamical systems model of how asymmetries in communication channel preferences contribute to the double empathy problem by generating vulnerabilities in cross-neurotype interactions. The model extensively analyzes how communication mismatch drives the interaction cost in the channel preference dimension of dyadic emotional bonding. It represents a promising first step in the overarching quantitative understanding of how internal psychological states modulate and are modulated by social exchanges that neuropsychiatry has begun to develop. 

The main theoretical contribution of our framework is to demonstrate a formal mechanism for how relative asymmetry in empathic signal production and empathic signal perception and their preservation under stress can significantly contribute to downward trajectories of mutual interaction and outcomes. While acknowledging that communication profiles are not universal within neurological types, the application of this general diagnosis-agnostic framework to the canonical channel preferences of neurotypical and autistic individuals replicates the negative trajectory of cross-neurotype interactions outlined by the double empathy problem. 

Under conditional experimental validation, this model shows the possibility of identifying specific therapeutic targets and interventions. Merely analyzing the proposed dynamical equations for the verbal-nonverbal dimension of communication mismatch and the double empathy problem, we were able to identify a possible mechanism for mitigating mutual empathy degradation. The most direct implication of the present formalization is that stability may be improved by preserving channels that remain informative to the partner. As the process of formalization evolves in future work, this research should focus on identifying the sources of cost for both partners, increasing mutual legibility, supporting explicit repair, and distributing accommodation reciprocally, without targeting individual behavioral modulation or requiring the autistic individual to approximate non-autistic communicative norms. 

The analysis presented here outlines a clear path of future theoretical research for the study of cross-neurotype interaction and emotional bonding. The immediate continuation would involve incorporating the dynamics of positive empathic trajectories that generalize beyond the focus on collapse or degradation. Afterwards, the efforts should focus on unifying into a single formal account multiple dimensions of dyadic interaction other than channel preference, such as other forms of communicative mismatch, affective resonance, and social-emotional motivation. The goal should be to assemble an empirically validated theoretical construct of how the internal state of two individuals determine their interaction outcomes and how these dynamics are perturbed by the consideration of diagnosis.  

The double empathy problem has been characterized as a bidirectional difficulty rather than a deficit residing in one party. The present model formalizes how this insight applies to the verbal/nonverbal dimension of dyadic interaction and how this is relevant for cross-neurotype communication when the median communication profiles of neurotypical and autistic individuals are used. The perception and output asymmetries built into the model produce a theoretical system where small differences in individual characteristics (particularly the output channel profile under stress) can determine whether an interaction remains stable or cascades toward mutual withdrawal. It represents a valuable starting point toward a mathematical formalization of all processes underpinning cognitive and emotional costs in dyadic interactions and how these mechanisms are altered by neurological type. 

\backmatter

\bmhead{Acknowledgements}
The authors are grateful to Cíntia V. C. Heidemann for several extremely useful comments made during the course of the research presented here.

\section*{Declarations}

\subsection*{Funding}
Not applicable.

\subsection*{Competing interests}
The authors have no competing interests to declare.

\subsection*{Ethics approval and consent to participate}
Not applicable.

\subsection*{Consent for publication}
Not applicable.

\subsection*{Data availability}
The data files of the input parameters and the outputs of the simulations are available from the corresponding author upon reasonable request.

\subsection*{Materials availability}
Not applicable. 

\subsection*{Code availability}
The codes used for the simulation, for computing the separatrix, and for plotting Figures 4 and 5 are available from the corresponding author upon reasonable request.

\subsection*{Author contributions}
Conceptualization and Methodology: all authors. Software, Formal analysis, Visualization, Writing — original draft: E.C. Writing — review and editing: all authors. Supervision: T.F., M.C.V., F.K.

\subsection*{Declaration of Generative AI and AI-assisted technologies in the writing process}
During the preparation of this work the authors used ChatGPT and Claude to improve the language and readability of the manuscript and to assist with manuscript and reference formatting. After using these tools, the authors reviewed and edited the content as needed and take full responsibility for the content of the publication.

\bibliography{sn-bibliography}
\end{document}